\documentclass[journal=langd5]{achemso}

\usepackage{graphicx}
\usepackage{amsmath}
\usepackage{amssymb}
\usepackage{multirow}
\usepackage{booktabs}
\usepackage[T1]{fontenc}
\usepackage{bm}
\SectionsOn
\setkeys{acs}{articletitle = true,keywords = true}

\usepackage[normalem]{ulem}
\usepackage[usenames,dvipsnames]{xcolor} 

\usepackage{titlesec}

\titleformat*{\section}{\large\bfseries}
\titleformat*{\subsection}{\normalsize\bfseries}

\SectionNumbersOff

\newif\ifRPSHighlitedChanges
\def\ifRPSHighlitedChanges{\iftrue}
\def\ifRPSHighlitedChanges{\iffalse}
\ifRPSHighlitedChanges
  \def\IRPS#1{{\color{blue}#1}}                       
  \def\SRPS#1{{\color{blue}\sout{#1}}}                
  \def\CRPS#1{{\color{blue}\textbf{(Comment: #1)}}}   
\else
  \def\IRPS#1{#1}
  \def\SRPS#1{\relax}
  
  \def\CRPS#1{\relax}
\fi

\newif\ifDKHighlitedChanges
\def\ifDKHighlitedChanges{\iftrue}
\def\ifDKHighlitedChanges{\iffalse}
\ifDKHighlitedChanges
  \def\EDITSDK#1{{\color{blue}#1}}
  \def\STRIKEDK#1{{\color{blue}\sout{#1}}}
  
\else
  \def\EDITSDK#1{#1}
  \def\STRIKEDK#1{\relax}
\fi

\newif\ifNeedCitation
\def\ifNeedCitation{\iftrue}
\ifNeedCitation
  \def\needCitation#1{{\color{red}\textbf{(Need Citation)}}}
\else
\fi

\newif\ifCheckNeeded
\def\ifCheckNeeded{\iftrue}
\ifCheckNeeded
  \def\checkNeeded#1{{\color{blue}\textbf{(Need Citation)}}}
\else
\fi

\title{Bridging Heterogeneity Dictates the Microstructure and Yielding Response of Polymer-Linked Emulsions}

\author{Daniel P. Keane}
\affiliation{Department of Chemical Engineering, University of Rhode Island, Kingston, RI 02881}

\author{Colby J. Constantine}
\affiliation{Department of Chemical Engineering, University of Rhode Island, Kingston, RI 02881}

\author{Matthew D. Mellor}
\affiliation{Department of Chemical Engineering, University of Rhode Island, Kingston, RI 02881}

\author{Ryan Poling-Skutvik}
\email{ryanps@uri.edu}
\affiliation{Department of Chemical Engineering, University of Rhode Island, Kingston, RI 02881}

\date{\today}

\begin{document}
	
%
%

\maketitle

\begin{abstract}
Soft materials possessing tunable rheological properties are desirable in applications ranging from 3D printing to biological scaffolds. Here, we use a telechelic, triblock copolymer polystyrene-\textit{b}-poly(ethylene oxide)-\textit{b}-polystyrene (SEOS) to form elastic networks of polymer-linked droplets in cyclohexane-in-water emulsions. The SEOS endblocks partition into the dispersed cyclohexane droplets while the midblocks remain in the aqueous continuous phase, resulting in each chain taking on either a looping or bridging conformation. By controlling the fraction of chains that form bridges, we tune the linear elasticity of the emulsions and generate a finite yield stress. Polymers with higher molecular weight ($\mathrm{M}_\mathrm{w}$) endblocks form stronger interdroplet connections and display a higher bridging density. Beyond modifying the linear rheology, the telechelic, triblock copolymers also alter the yielding behavior and processability of the linked emulsions. We examine the yield transition of these polymer-linked emulsions through large amplitude oscillatory shear (LAOS) and probe the emulsion structure through confocal microscopy, concluding that polymers that more readily form bridges generate a strongly percolated network, whereas those that are less prone to form bridges tend to produce networks composed of weakly-linked clusters of droplets. When yielded, the emulsions consisting of linked clusters break apart into individual clusters that can rearrange upon the application of further shear. By contrast, when the systems containing a more homogeneous bridging density are yielded, the system remains percolated but with a reduced elasticity and bridging density. The demonstrated ability of telechelic triblock copolymers to tune not only the linear viscoelasticity of complex fluids but also their nonlinear yield transition enables the use of these polymers as versatile and robust rheological modifiers. We expect our findings to therefore aid the design of the next generation of complex fluids and soft materials.
\end{abstract}

\section{Introduction}

Triblock copolymers possess a unique structure consisting of three distinct polymer blocks of varying chemistries, which promotes their self-assembly into complex structures in both melts and suspensions. In melts of triblock copolymers, the relative size of the blocks controls pattern formation and other complex phase behaviors.\cite{Matsen1994, Bailey2002, Epps2004} In solution, the assembly of telechelic polymers depends strongly on the polymer concentration and solvent quality. At low polymer concentrations in a midblock-selective solvent, the endgroups of a telechelic polymer localize to form flower-like micelles \cite{Ganguly2005,Lu2015,Huo2022,Silva2022} or more complex structures such as vesicles.\cite{Nardin2000} As the polymer concentration is increased, the midblocks begin to bridge endblock-rich domains to form networks of physically cross-linked micelles.\cite{Lemmers2011} The endblocks of telechelic polymers localize into micelles or droplets of emulsions with each telechelic linker preferentially forming loops, such that the endblocks partition into the same droplet or micelle, or bridges, such that each endblock is embedded in a different droplet or micelle. Other conformations such as dangling or free chains are not energetically favorable, especially for higher molecular weight ($\mathrm{M}_\mathrm{w}$) species, and are not expected to be taken by an appreciable fraction of chains.\cite{Testard2008} The unique structure and self-assembly characteristics of these polymers allow them to serve as functional rheological modifiers in dispersed systems such as colloidal suspensions and emulsions.\cite{Lesueur1998,Xia2017,Xia2019} 

The ability to control the rheological properties of suspensions is vital to their performance in applications ranging from cosmetics\cite{Brummer1999,Venkataramani2020,Bezerra2021} to pharmaceuticals\cite{Matos2018, Light2021}, and 3D printing \cite{Sommer2017} to injection molding.\cite{Moon2020} The rheological properties of neat emulsions are largely controlled by the volume fraction of the dispersed phase\cite{Einstein1906,Taylor1932,Oldroyd1955,Pal2000} $\phi$ as well as the size and elasticity of the dispersed droplets.\cite{Derkach2009,Shu2013,Poling-Skutvik2020,Sanatkaran2021} When $\phi$ is raised above the random close packing of spheres $\phi_c$, droplets deform as the surfaces come into contact, and the system takes on the form of a high internal phase emulsion. The contact points between droplets result in an elastic network with properties controlled by surface tension, droplet size, and continuous phase viscosity.\cite{Pal2006,Welch2006} To induce elasticity at lower volume fractions ($\phi<\phi_c)$, however, requires additives that generate attractive interactions between the droplets. These attractive interactions induce structural changes in the suspension, converting the randomly dispersed droplets into fractal aggregates that can support external stresses. Polymers can induce attractive interactions in suspensions through a variety of methods. Polymers may induce depletion interactions as a result of minimizing excluded volume, but the gels formed through this method tend to be slow-forming and brittle.\cite{Oosawa1954,Asakura1958,Meller1999,Lu2008} Alternatively, high-$\mathrm{M}_\mathrm{w}$ polymers can be used to thicken a suspension's continuous phase, but they do not fundamentally change the viscoelastic behavior of the emulsions.\cite{Pal1992} Finally, telechelic triblock copolymers with solvophobic endgroups can be added to emulsions to induce associative forces between droplets that can result in strongly elastic networks.\cite{Meng2006,Peters2016,Zinn2017,Keane2022,Larson2022}
 
 When used as a rheological modifier, the strength of the associative nodes formed by telechelic triblock copolymers is controlled by both the endblock association and enthalpic penalty of transferring the endblock to the continuous phase.\cite{Mishra2018} When the endgroups form micelles, endblock entanglement imposes a large energy penalty for pullout, but this behavior differs from that of physically cross-linked emulsions, in which the endblock entanglement is negligible. Instead, the strength of a telechelically linked networks of droplets is largely controlled by the thermodynamic penalty of transferring the endblock from the dispersed phase to the continuous. For low-$\mathrm{M}_\mathrm{w}$ endblocks, the energy penalty of chain pullout is low. Networks formed by these low-$\mathrm{M}_\mathrm{w}$ polymers tend to exist at equilibrium, with thermal fluctuations driving chains to transition between looping and bridging conformations, resulting in low moduli and viscous relaxations.\cite{MaloDeMolina2014,Puech2008} Increasing the size of the endblocks, however, generates permanent elasticity, as thermal fluctuations are no longer sufficient to drive chain rearrangements between droplets.\cite{Keane2022} As a result, the linked networks are kinetically trapped with static polymer conformations. It remains unclear, however, how the application of shear disrupts these kinetically trapped structures and how these structures evolve under stress.
 
In this work, we investigate how high-$\textrm{M}_\textrm{w}$ telechelic copolymers affect the yield transitions and modify the bridging density in emulsions. We utilize the telechelic, triblock copolymer polystyrene-\textit{b}-poly(ethylene oxide)-\textit{b}-polystyrene (SEOS) of varying $\textrm{M}_\textrm{w}$ to physically crosslink cyclohexane-in-water emulsion droplets. The hydrophobic polystyrene (PS) endblocks of SEOS partition into the cyclohexane droplets, and the hydrophilic poly(ethylene oxide) (PEO) midblocks remain in the continuous phase. We examine the yield transition and response to processing of the polymer-linked emulsions through large amplitude oscillatory shear (LAOS) rheology. We relate the rheological curves to specific yielding mechanisms and explore structural changes that occur when the network is fractured. Developing a better understanding of how these systems evolve under shear will assist in the design of novel materials for a range of applications, including 3D printing\cite{Sommer2017} and biological scaffolds.\cite{Lin2013,Aguilar-De-leyva2020}

\section{Materials and Methods}
\subsection{Materials}

Cyclohexane ($\geq$99\% purity), glycerol, Tween 20, Span 20, and 9,10-bis(phenylethynyl)-anthracene were purchased from Sigma Aldrich. Polystyrene-\textit{b}-poly(ethylene oxide)-\textit{b}-polystyrene (SEOS) of different molecular weights were purchased from PolymerSource and used without further modification. Linear poly(ethylene oxide) (PEO) was purchased from Scientific Polymer. Polymer nomenclature and polymer molecular weights are presented in Table \ref{tab:SEOS}. Each polymer is named according to SEOS-X/Y, where X and Y denote the number average molecular weight $\mathrm{M}_\mathrm{n}$ in kDa of the midblock and endblock, respectively. 

\begin{center}
\begin{table}
\begin{tabular}{|c c c|} 
 \hline
 \multirow{2}{*}{Polymer}&$\mathrm{M}_{\mathrm{n}}$ [kDa] & \multirow{2}{*}{\DJ} \\ [0.1ex]
 &(PS-\textit{b}-PEO-\textit{b}-PS)&\\
 \hline
  SEOS-180/32 & 32-\textit{b}-180-\textit{b}-32 & 1.10 \\
 \hline
 SEOS-180/10 & 10-\textit{b}-180-\textit{b}-10 & 1.10 \\
 \hline
 SEOS-27/11.5 & 11.5-\textit{b}-27-\textit{b}-11.5 & 1.09 \\
 \hline
  PEO & 0-\textit{b}-222.3-\textit{b}-0 & 1.15 \\
 \hline
\end{tabular}
\caption{\label{tab:SEOS} Nomenclature, number-average molecular weights $\mathrm{M}_\mathrm{n}$, and dispersities \DJ{} of polymers used in this study.}
\end{table}
\end{center}

\subsection{Emulsion Preparation}
Emulsions were prepared as described in our previous work.\cite{Keane2022} Briefly, a nonionic surfactant mixture of Span 20 (Hydrophilic-Lipophilic Balance (HLB) $=8.6$) and Tween 20 (HLB $=16.7$) was prepared at a weight ratio of 1:4 to achieve an HLB of 15.0. This mixture was diluted with water to a ratio of 1:29 surfactant:water. Emulsions with volume fraction $\phi=0.5$ were prepared by adding 3 mL of cyclohexane dropwise to 3 mL of the aqueous surfactant solution at a rate of 0.3 mL/min while being sonicated by a Branson SLPe digital sonifier with a 1/8 in. tapered microtip. Sonication was continued for two minutes after the complete addition of cyclohexane to ensure full emulsification. The resulting emulsions contain a polydisperse distribution of droplets with an average hydrodynamic diameter of $d_H\approx 400$ nm, as determined through dynamic light scattering (DLS). Polymer was added at the desired concentration, the vials were wrapped with Parafilm, and the emulsions were stirred at 50 $^\circ$C until the polymer was fully dissolved. Polymer was added at nominal concentrations of 9.0, 27, and 46 g/L, which equate to 1, 3, and 5 wt\%, respectively. Emulsions remained stable for upwards of two weeks \EDITSDK{as confirmed by constant droplet size distributions as determined through DLS and rheological properties.} All emulsion characterization was conducted within this time frame.

To prepare index-matched emulsions, the same procedure as above was used with a few minor alterations. A fluorescent dye, 9,10-bis(phenylethynyl)anthracene (BPA), was added to cyclohexane at a concentration of $\sim$1 g/L prior to emulsification. BPA has excitation and emission wavelengths of $\lambda_\mathrm{ex}=313$ nm and $\lambda_\mathrm{em}=491$ nm, respectively.\cite{IsadoreBBerlman1971} After emulsification, the vials were wrapped with aluminum foil to prevent photobleaching, and 9 g/L polymer was added. When the polymer was fully dissolved, glycerol was added to bring the continuous phase to 68.5\% glycerol by weight. This aqueous glycerol solution has a refractive index of 1.42, matching that of cyclohexane.\cite{Aminabhavi1996} The resulting emulsion was optically clear (Fig. S3), indicating successful index matching. The final volume fraction $\phi$ of cyclohexane and polymer concentration $c$ in the index-matched systems are approximately 0.24 and 4.3 g/L, respectively.

\subsection{Rheology}
Rheology was conducted on a TA Instruments HR-20 rheometer using 2$^\circ$ cone-and-plate geometries with diameters of 20 and 60 mm\EDITSDK{, each with a minimum gap size of 54 $\mu$m}. A solvent trap covered the sample and was sealed by mineral oil. We added $\sim$2 mL cyclohexane to the interior of the solvent trap to minimize sample evaporation. Samples remained stable to evaporation for several hours, as shown in Fig. S2. 

The primary tests used to examine the rheological properties of the materials in this study were amplitude sweeps conducted from strains $\gamma=$0.01 to 100\% and then in the reverse direction, with five data points collected per decade. Tests were conducted at oscillation frequencies ranging from $\omega = 0.1-100$ rad/s, and a minimum of three trials were conducted for each pairing of polymer $c$, polymer $\mathrm{M}_\mathrm{w}$, and $\omega$. Prior to collecting data, the sample was equilibrated for 180 s at $T = 25^\circ$ C to reach steady state. Following this equilibration step, the sample was subjected to a five-minute oscillation at $\omega=$10 rad/s and $\gamma=$0.1\% strain to attempt to remove loading effects.

\subsection{Confocal Microscopy}
Confocal microscopy was conducted on a Nikon Eclipse Ti2 Inverted Confocal Microscope. Samples were imaged in sealed chambers, crafted from glass slides and sealed with epoxy to minimize evaporation and bulk flow when viewing the samples under the microscope. All images were captured using a 40$\times$ objective (pixel size = 0.31 $\mu$m)  and a 488 nm green filter. Image collection was conducted by a 5$\times$5 raster over an area of 25 mm$^2$ such that collected images were representative of a random sampling and no location was subjected to an extended exposure to the laser, minimizing photobleaching effects. Each image was collected with an exposure time of 10-20 seconds.

Before image analysis, each image was passed through an image processing algorithm to remove noise and convert them to 8-bit. The images were denoised using a deep neural network (\textit{DnCNN}) and then binarized based on a global brightness threshold. Holes of four pixels or less were then closed using MATLAB's \textit{imclose} function and isolated groupings of three or less pixels were removed with the \textit{imopen} function. This image processing was conducted using MATLAB's Image Processing Toolbox as shown in Fig. S4. The collected images were then analyzed with MATLAB's \textit{regionprops} function to determine the size of droplets and clusters.

\section{Results and Discussion}

\subsection{A Review of Linear Behavior}

\begin{figure*}[h!]
\includegraphics[width = 6.5 in]{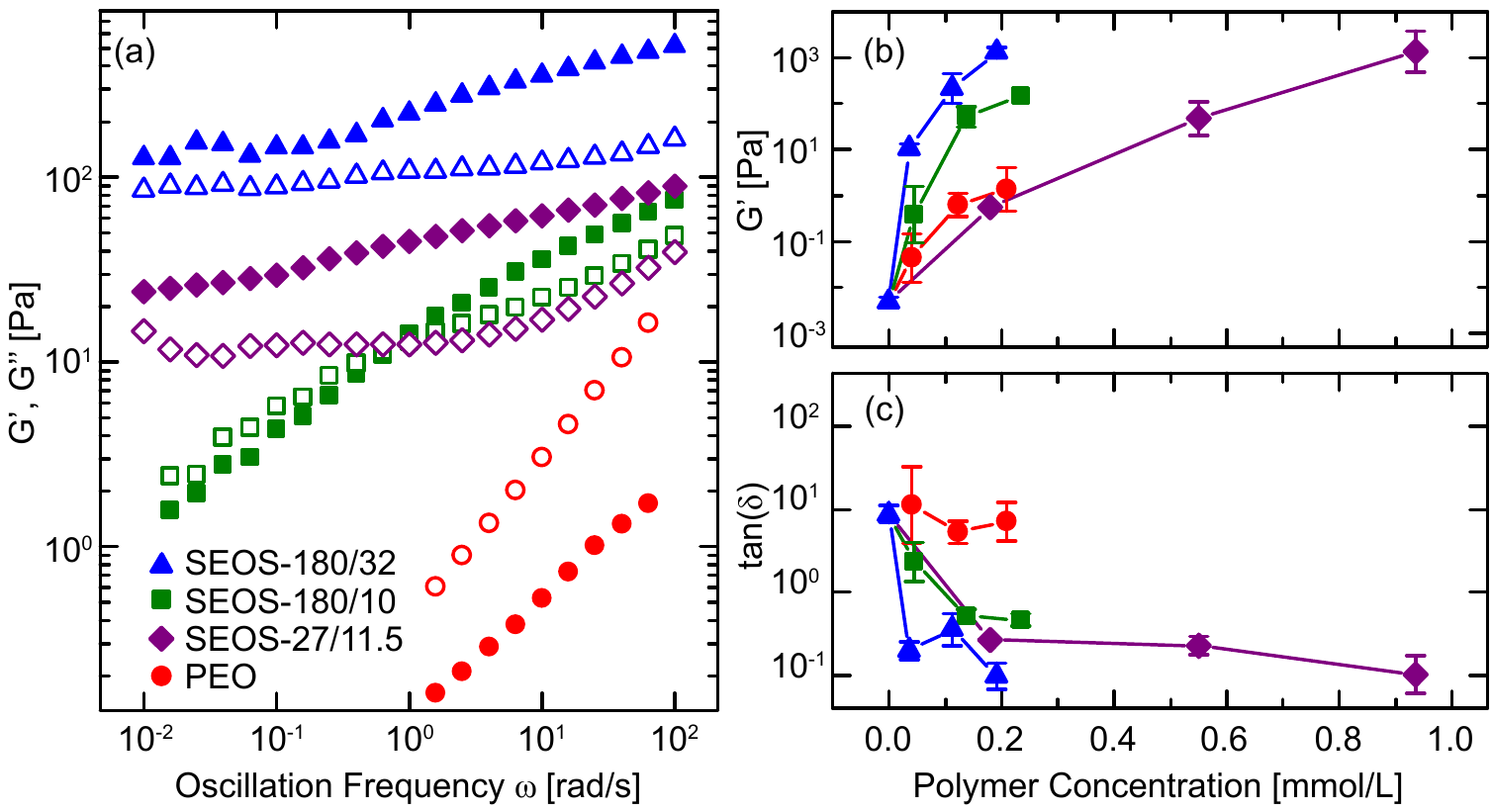}
\caption{\label{fig:Linear} (a) Storage modulus G' (closed) and loss modulus G'' (open) as a function of oscillation frequency $\omega$ for emulsions containing 27 g/L of the specified polymer\EDITSDK{, corresponding to molar concentrations of 0.11, 0.14, 0.55, and 0.12 mmol/L for SEOS-180/32, SEOS-180/10, SEOS-27/11.5, and PEO, respectively}. (b) G' and (c) $\mathrm{tan}(\delta) = \mathrm{G''/G'}$ at $\omega = 10$ rad s$^{-1}$ as a function of polymer concentration. Error bars represent one standard deviation \IRPS{measured across samples}. Figure adapted from Keane, D. P.; Mellor, M. D.; Poling-Skutvik, R. Responsive Telechelic Block Copolymers for Enhancing the Elasticity of Nanoemulsions. \textit{ACS Appl. Nano. Mater.} \textbf{2022}, \textit{5}, 5934-5943. Copyright 2022 American Chemical Society.}
\end{figure*}

Adding telechelic triblock copolymers to suspensions of emulsion droplets results in the formation of polymer bridges that link droplets together, significantly increasing the suspension elasticity and slowing their relaxations. In our previous work\cite{Keane2022}, we investigated how varying polymer molecular weight $\mathrm{M}_\mathrm{w}$ and polymer concentration $c$ affect the polymer bridging density and suspension linear rheology (Fig. \ref{fig:Linear}). The telechelic SEOS polymers increase the stiffness and elasticity of the emulsions far more than the addition of hydrophilic PEO, which is incapable of forming the same interdroplet connections. This increase in the mechanical properties is the result of an increase in the bridging density, or number of bridging chains $n_\mathrm{bridging}$, between droplets and the energy penalty of chain pullout $E_\mathrm{pullout}$. The likelihood of chains forming bridges scales with the $\mathrm{M}_\mathrm{w}$ of the midblock\cite{Testard2008,Puech2008}, as the entropic penalty of stretching enough to bridge droplets is greater for the polymers with smaller $\mathrm{M}_\mathrm{w}$ midblocks.\cite{Rubinstein2003} Therefore, $n_\mathrm{bridging}$ increases with $c$ and midblock $\mathrm{M}_\mathrm{w}$. Chains with a 180 kDa midblock are much more likely to form bridges than those with a 27 kDa midblock, leading to a more rapid increase in suspension elasticity with polymer concentration for the high-$\mathrm{M}_\mathrm{w}$ SEOS-180/10 and SEOS-180/32 than for the low-$\mathrm{M}_\mathrm{w}$ SEOS-27/11.5. Furthermore, the size of the polymer endblock controls $E_\mathrm{pullout}$, which is proportional to $\Delta \chi N_\mathrm{PS}$, where $N_\mathrm{PS}$ is the size of the PS endblock, and $\Delta \chi$ is the difference in interaction parameters of polystyrene/cyclohexane ($\chi_\mathrm{PS/CH}$) and polystyrene/water ($\chi_\mathrm{PS/H_2O}$). We therefore observe higher elasticities in emulsions containing SEOS-180/32 than in those containing the same $c$ of SEOS-180/10. Based on these linear properties, we conclude that emulsions containing SEOS-180/32 have stronger interdroplet interactions than the other SEOS species and that a greater fraction of chains are in the bridging conformation for SEOS-180/32 and SEOS-180/10 than for SEOS-27/11.5. \IRPS{Additional details on the linear rheology of these polymer-linked emulsions is included in Ref. \citenum{Keane2022}}. Although we have shown that the polymer $\mathrm{M}_\mathrm{w}$ controls the bridging fraction $\epsilon=n_\mathrm{bridging}/n_\mathrm{total}$ and the strength of interparticle interactions, it remains poorly understood how these multicomponent complex fluids yield under applied stress, how the polymer affects the yield stress, and by what physical mechanisms the systems yield.

\subsection{Nonlinear Rheology}
\begin{figure}[ht]
\includegraphics[width = 3.25 in]{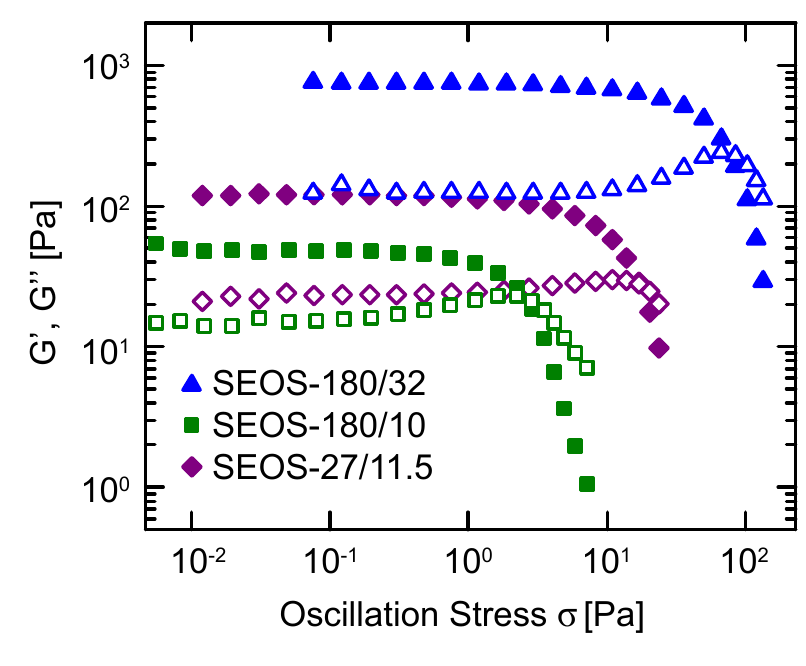}
\caption{\label{fig:AmpSweep1} Storage modulus G' (closed) and loss modulus G'' (open) measured at $\omega=$ 10 rad s$^{-1}$ as a function of oscillation stress $\sigma$ for emulsions containing 27 g/L of the specified polymer.}
\end{figure}

We now investigate how these differences in the bridging density and emulsion elasticity affect the yielding mechanism in these linked emulsions. Although the yield transition can be investigated through a variety of rheological procedures,\cite{Ewoldt2013,Fernandes2017,Lee2017,Donley2020} here we will primarily use oscillatory amplitude sweeps in which we subject the sample to oscillations at a frequency $\omega = 10$ rad/s and vary the amplitude strain. By probing the nonlinear rheology through LAOS, we can observe the evolution of the storage and loss moduli across the yield transition and identify how the telechelic triblock copolymers contribute to energy dissipation and structural changes in the linked emulsions. Because the yield transition is typically associated with a particular stress, we plot the viscoelastic moduli as a function of the applied stress on the sample (Fig. \ref{fig:AmpSweep1}). 

At low stresses $\sigma$, we observe a linear viscoelastic region (LVR) for all polymer-linked emulsions, indicated by a plateau in G' and G''. Because these plateaus correspond to the linear response of the system, they follow the same trends as shown in Fig.\ \ref{fig:Linear}, with higher $\mathrm{M}_\mathrm{w}$ mid- and endblocks leading to higher moduli. As $\sigma$ increases, the samples yield and transition from elastically dominated (i.e., G' > G'') to viscously dominated (i.e., G'' > G') regimes. During this yield transition, G' initially decreases gradually and then sharply as the elastic network formed by polymer-bridged droplets is broken. By contrast, G'' exhibits either an overshoot or an immediate decay depending on the experimental conditions and sample composition. The decay in G'' is more gradual than that of G', leading to a crossover at a stress $\sigma_\mathrm{y}$ corresponding to the yield stress of the system. We note that although the definition of the yield stress is ambiguous\cite{Dinkgreve2016,Malkin2017} with some articles defining it as the end of the LVR or the stress at the peak of the G'' overshoot, the crossover is G' and G'' is a convenient metric for these samples. \EDITSDK{We chose this definition because not all samples display an overshoot in G'' and the end of the LVR can be challenging to clearly identify due to subtle changes in the linearity of the of G'' even at low very low stresses.} The presence of an overshoot in G'' is generally attributed to the conversion of elastic elements in the material to fluid-like energy dissipation as irreversible plastic deformation starts to dominate.\cite{Donley2020} We observe some qualitative differences in the yield transition for emulsions containing the different telechelic linkers. First, samples with larger moduli tend to yield at a higher stress, consistent with trends observed for other viscoelastic materials.\cite{Cubuk2017} Second, the prominence of the G'' overshoot increases with increasing moduli, which suggests that yielding stiffer emulsions converts more elastic elements into viscous dissipation. As the moduli depend on the $\mathrm{M}_\mathrm{w}$ of the contained polymer, these trends further suggest that the yielding of the emulsions containing SEOS-180/32 occurs at a greater stress and dissipates more energy than that of emulsions containing the lower $\mathrm{M}_\mathrm{w}$ polymers. To quantify these trends, we investigate their dependence on the frequency at which the sample is yielded, the polymer concentration $c$, and the polymer $\mathrm{M}_\mathrm{w}$.

\subsection{Frequency Effects}
\begin{figure}[tb!]
\includegraphics[width = 3.25 in]{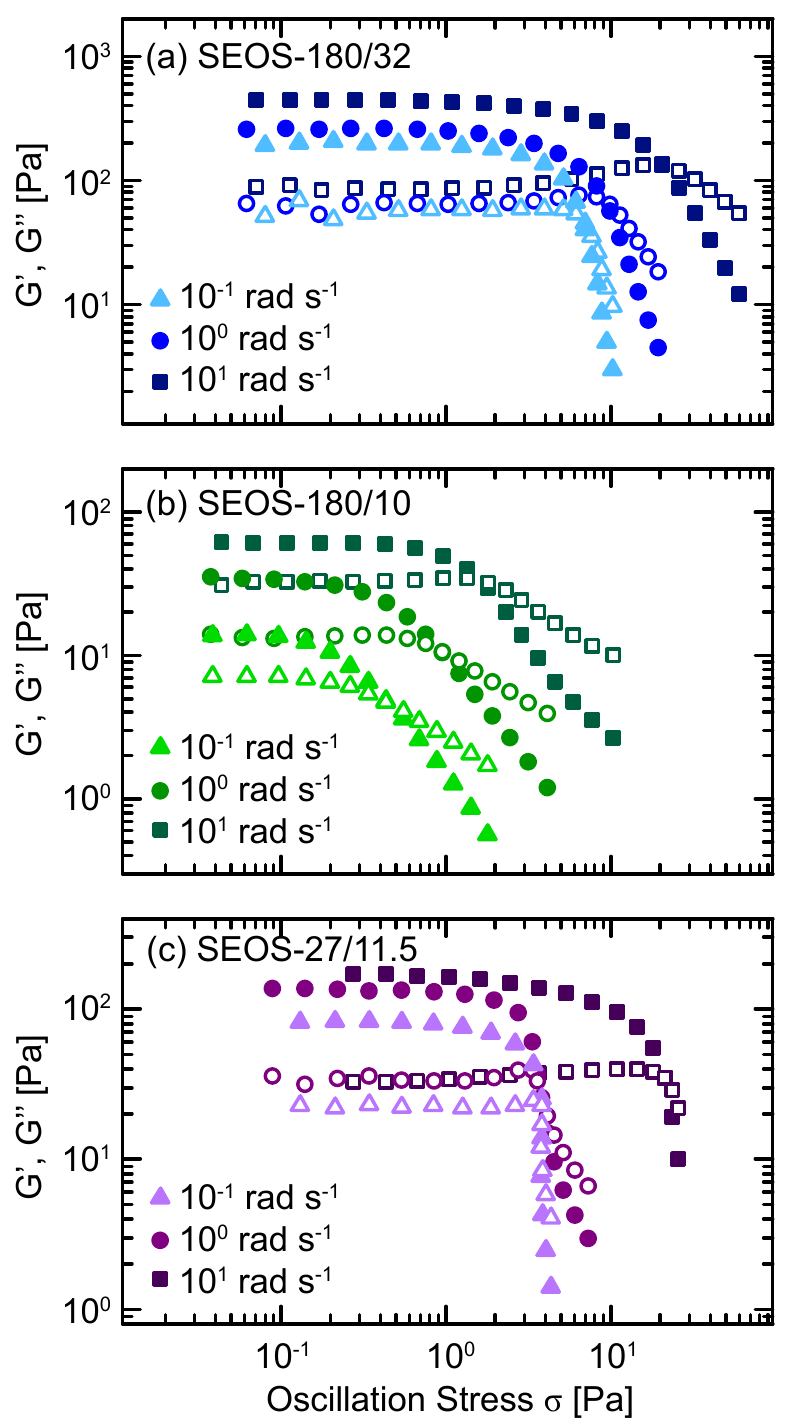}
\caption{\label{fig:AmpSweeps} Storage modulus G' (closed) and loss modulus G'' (open) as a function of oscillation stress $\sigma$ for emulsions containing 27 g/L (a) SEOS-180/32, (b) SEOS-180/10, and (c) SEOS-27/11.5 measured at various frequencies.}
\end{figure}

As observed in Fig. \ref{fig:Linear}, the choice of polymer significantly affects the relaxation behavior and frequency dependence of systems, with high-$\mathrm{M}_\mathrm{w}$ endblocks exhibiting permanent elasticity and lower $\mathrm{M}_\mathrm{w}$ linkers allowing viscous flow on long time scales. Based on the frequency effects observed in the linear regime, we hypothesize that the yielding of the droplet network may similarly depend on frequency. To test this hypothesis, we conduct oscillatory amplitude sweeps at different frequencies $\omega$ (Fig. \ref{fig:AmpSweeps}). In agreement with the linear measurements, the moduli increase with frequency regardless of the contained polymer. Furthermore, emulsions displaying a higher elastic modulus in the LVR generally yield at higher stresses, meaning that the yield stress generally increases with $\omega$ as well. In addition to $\sigma_\mathrm{y}$, both the height of the G'' overshoot and the decay rate following yielding appear to depend on polymer $M_\mathrm{w}$ and $c$. For emulsions containing SEOS-180/32, we observe sharper decays in G' and less pronounced G'' overshoots at low $\omega$. Emulsions containing SEOS-27/11.5 exhibit similar trends, but they are less pronounced than in the SEOS-180/32 emulsions. Finally, emulsions containing SEOS-180/10 exhibit differences in the moduli and stress at which the materials yield, but the overall shape is constant for all $\omega$. These contrasting frequency dependencies suggest that the polymer-linked emulsions possess distinct network architectures, arising from the different bridging density and interaction strength of the contained telechelic polymers.

\begin{figure}[tb!]
\includegraphics[width = 3.25 in]{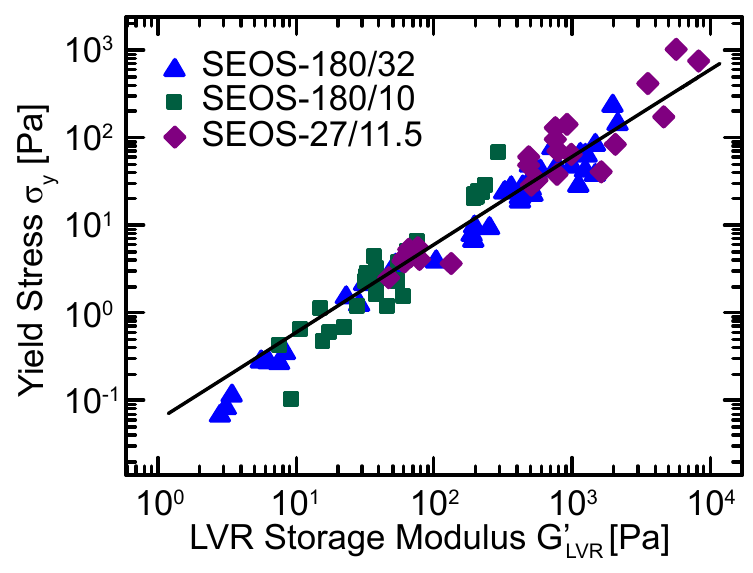}
\caption{\label{fig:YieldStress} Yield stress $\sigma_y$ as a function of the storage modulus $\mathrm{G'}_\mathrm{LVR}$ in the LVR. Displayed points are for all tested polymer concentrations and oscillation frequencies $\omega$. The data is fitted to the equation $\mathrm{G'}_\mathrm{LVR}=0.06\sigma_y$.}
\end{figure}

To further investigate these differences, we quantify three key characteristics: (i) the yield stress $\sigma_y$, identified as the stress at the crossover of G' and G''; (ii) the height of the G'' overshoot, quantified by the ratio of the maximum of G'' relative to the LVR value ($\textrm{G''}_\textrm{max}/\textrm{G''}_\textrm{LVR}$); and (iii) the exponent $\beta$ characterizing the maximum decay rate of G' according to \EDITSDK{$\mathrm{G'}=A\sigma^{-\beta}$}. These parameters are shown schematically in Fig. S7. These metrics have been used in earlier work to identify the mechanisms and physics underlying the yielding of soft materials.\cite{Urayama2014,Corker2019,Fan2019,Morlet-Decarnin2022,Rathinaraj2022,Sudreau2022} We first consider $\sigma_y$. Emulsions for which $\textrm{G''}>\textrm{G'}$ in the LVR are fluid-like and therefore do not display a crossover in G' and G'', so these trials are omitted from this analysis. We observe that $\sigma_y$ spans six orders of magnitude depending on polymer M$_\mathrm{w}$ and $c$. Moreover, $\sigma_y$ scales linearly with $\textrm{G'}_\textrm{LVR}$ (Fig. \ref{fig:YieldStress}) across all $\omega$, $\textrm{M}_\textrm{w}$, and $c$ with the data following $\sigma_y=0.06\mathrm{G'}_\mathrm{LVR}$. This slope indicates a yield strain of $\gamma_y=6.0\% \pm 0.3\%$, consistent with the yielding of many classes of amorphous solids\cite{Cubuk2017}. This linear relationship suggests that yielding in these polymer-linked emulsions occurs primarily via the rearrangement of droplets. Although viscous dissipation may occur through droplet rearrangement, the polymer properties control the stress necessary to induce these arrangements. Thus, we seek to better understand the polymer effects on the yielding through the other metrics that we have identified.

\begin{figure*}[ht]
\includegraphics[width = 6.5 in]{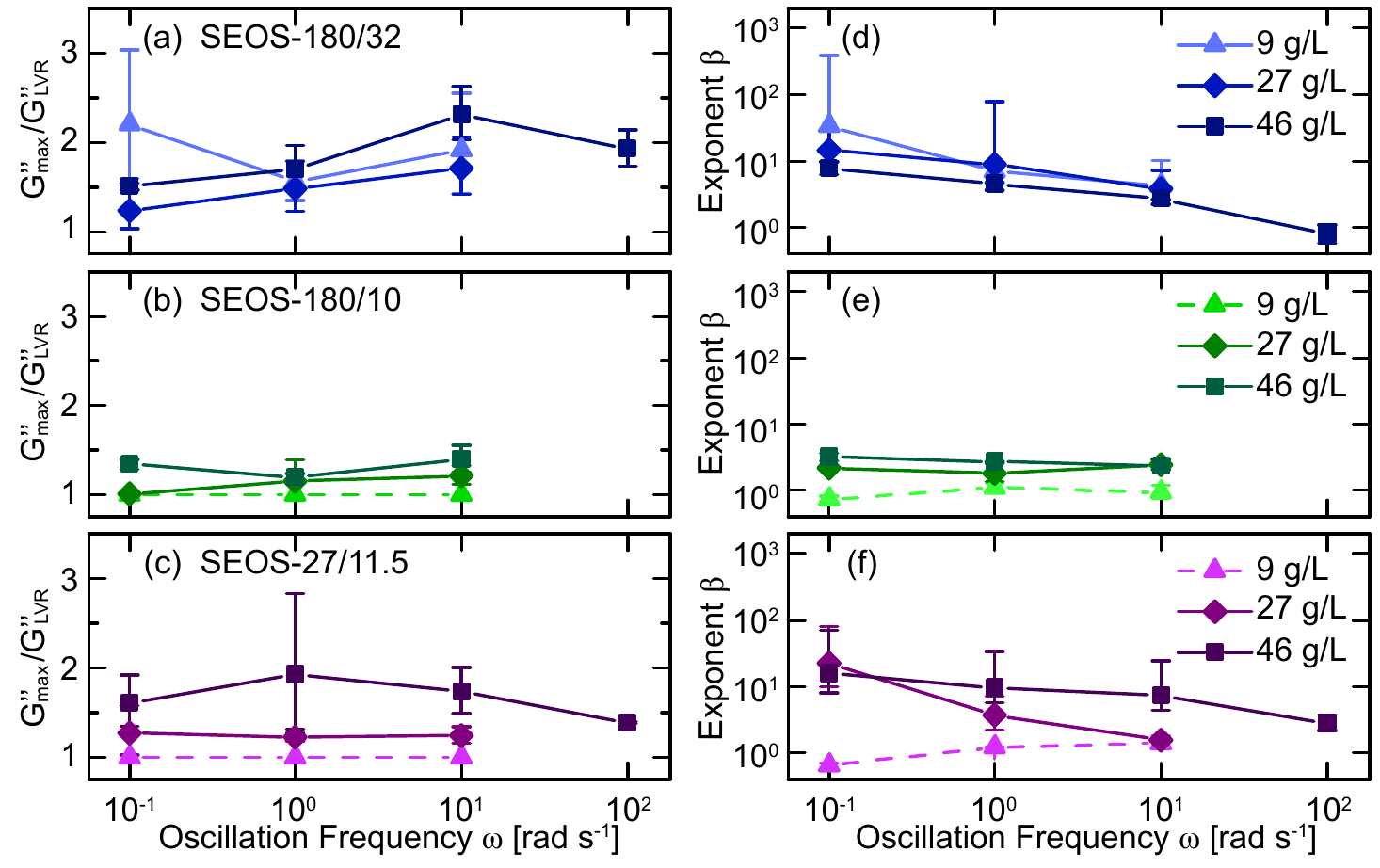}
\caption{\label{fig:slopesAndHeights} (a-c) $\textrm{G''}_\textrm{max}$/$\textrm{G''}_\mathrm{LVR}$ and (d-f) power law exponent $\beta$ as a function of oscillation frequency $\omega$ for emulsions containing varying concentrations of (a, d) SEOS-180/32, (b, e) SEOS-180/10, and (c, f) SEOS-27/11.5. Dashed lines indicate samples that did not show a discernible overshoot in G''. Error bars show one standard deviation \EDITSDK{measured across samples}.}
\end{figure*}

\begin{figure}[tb!]
\includegraphics[width = 3.25 in]{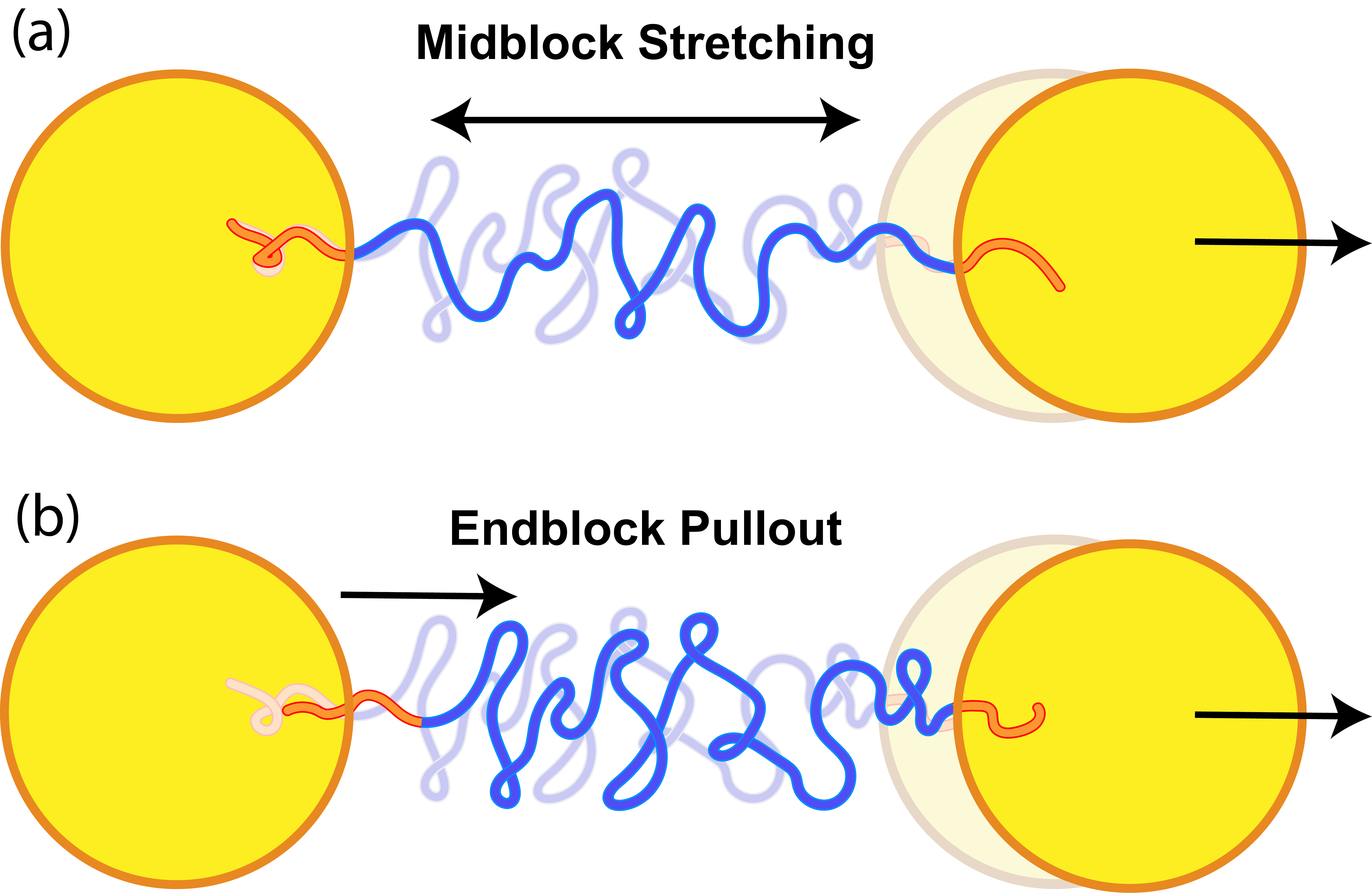}
\caption{\label{fig:Stretching} Schematic depicting the two processes that occur when two polymer-linked droplets are pulled apart: (a) midblock stretching which stores energy elastically, and (b) partial endblock pullout which dissipates energy.}
\end{figure}

The next parameter that we consider is the height of the G'' overshoot (Fig. \ref{fig:slopesAndHeights}a-c). The G'' overshoot physically represents an increase in the rate of plastic energy dissipation before the material begins to flow, or in other words, the transition from solid-like energy storage to fluid-like energy dissipation.\cite{Parthasarathy1999,Donley2020,Morlet-Decarnin2022} The height of this overshoot should therefore relate to the energy required to deform the network under shear, which can occur via midblock stretching or endblock pullout (Fig. \ref{fig:Stretching}). The energy dissipated upon yielding, and therefore $\textrm{G''}_\textrm{max}/\textrm{G''}_\textrm{LVR}$, depend on \IRPS{three parameters: the number of chains experiencing pullout $n_\textrm{pullout}$, the energy penalty for each displaced endblock $E_\mathrm{pullout}$, and the elastic deformation of the midblock}. For all polymers \IRPS{and at all $\omega$}, the height of the G'' overshoot increases with polymer concentration within experimental error as there are more \IRPS{bridging} chains in the more concentrated systems. At the lowest concentrations of SEOS-180/10 and SEOS-27/11.5, no G'' overshoot is observed. We expect that bridges still experience pullout in these systems but that their contribution to energy dissipation is significantly less than that of droplet rearrangement, so the system does not display a detectable overshoot. Emulsions containing higher concentrations of SEOS-180/10 exhibit small overshoots in G'', indicating that yielding the system dissipates less energy through chain pullout than in emulsions containing SEOS-180/32 or SEOS-27/11.5. We attribute \EDITSDK{this decrease} in the height of the G'' overshoot to the weaker partitioning of low-$\mathrm{M}_\mathrm{w}$ endblocks, \IRPS{resulting in less energy dissipation upon yielding.}

\begin{figure}[tb!]
\includegraphics[width = 3.25 in]{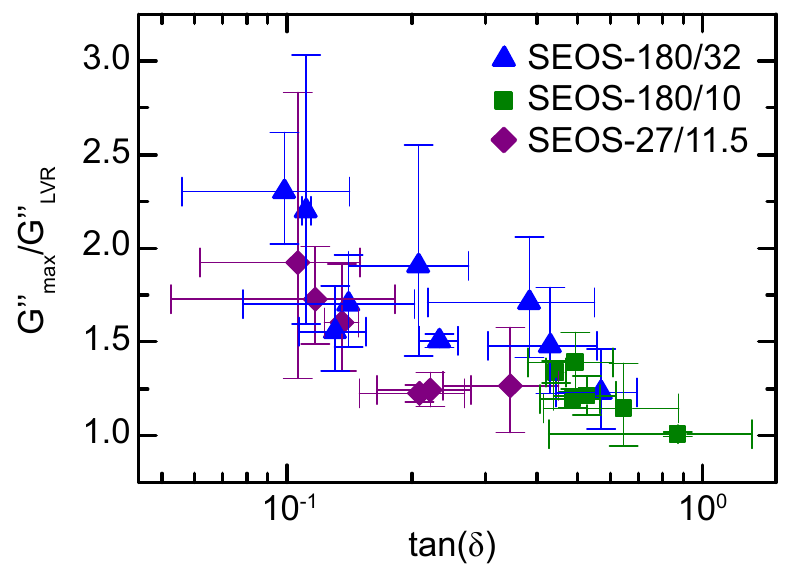}
\caption{\label{fig:tanDelta} \IRPS{The ratio of $\mathrm{G''_{max}/G''_{LVR}}$ as a function of tan($\delta$) at the same frequency within the LVR. Errors bars show one standard deviation between samples. }}
\end{figure}

\IRPS{The G'' overshoot increases with $\omega$ for the SEOS-180/32 samples but is largely independent for the SEOS-180/10 and SEOS-27/11.5 samples. To understand this dependence, we must consider the phenomena occurring as stress is applied to the polymer-linked emulsions. For small strains, short-range droplet rearrangements and chain stretching occur to accommodate the applied strain. When this strain is increased beyond the LVR, however, the network will plastically deform with chains undergoing the two simultaneous processes demonstrated in Fig. \ref{fig:Stretching}. The polymer midblocks stretch entropically, and the endblocks pull out from the droplet interiors. When the energy required to stretch the midblock exceeds the enthalpic penalty of transferring the PS endblock to the continuous phase, the endblock is forced into the aqueous phase, eventually re-embedding into a droplet to generate a new bridge or loop. With this physical picture, we can attribute the differing frequency dependencies in the size of the G'' overshoot to the segmental relaxations of the midblock. When the measurement frequency is faster than the relaxation of the midblock, the polymer chains have insufficient time to relax. Therefore, when chain pullout occurs, the dissipated energy is a sum of both the enthalpic penalty of displacing the endblock into the continuous phase as well as the entropic stretching of the midblock. Alternatively, at low frequencies, chains can partially relax during the yielding such that the energy released upon pullout is attributable to only the endblock displacement. From this physical picture, we can understand why a frequency dependence is only observed for the SEOS-180/32 sample, which exhibits an increase in the linear elasticity of the SEOS-180/32 emulsions for $\omega \gtrsim 1$ rad s$^{-1}$ (Fig. \ref{fig:Linear}). Finally, we account for all three effects -- bridging density, $E_\mathrm{pullout}$, and segmental relaxations -- by collapsing the magnitude of the G'' overshoot as a function of $\tan(\delta)$ (Fig. \ref{fig:tanDelta}). This parameter quantifies the relative elasticity of these emulsions, demonstrating that increasingly elastic systems (i.e., smaller $\tan(\delta)$) exhibit larger overshoots in G'' as elastic deformation is converted into viscous dissipation.}

Based on these findings, we conclude that yielding occurs as the material transitions from storing energy elastically through the stretching of bridged polymers to dissipating energy through chain pullout and subsequent droplet rearrangement. With this mechanism, the rate of decrease in elasticity upon yielding should reflect the distribution of stresses that the bridged polymers can support. We quantify the steepness of the decay in the elastic modulus G' as a function of stress according to a power-law $G'(\sigma)=A\sigma^{-\beta}$ where $\beta$ captures the maximum decay rate of G' and $A$ is a mathematical prefactor (Fig. \ref{fig:slopesAndHeights}d-f). Due to instances of infinitely sharp decays, the reciprocal of the arithmetic mean of the reciprocals\cite{Ferger1931} was used in place of an arithmetic mean. Steeper slopes, or larger values of $\beta$, indicate a sharper yielding with chains undergoing pullout over a narrow range of $\sigma$. Alternatively, a low $\beta$ value suggests that chain pullout is occurring over a wide range of $\sigma$. If bridging chains were distributed uniformly throughout the sample, the stress would be constant throughout the network so that all chains experience pullout simultaneously once $\sigma$ exceeds the thermodynamic penalty of chain pullout. By contrast, finite values of $\beta$ indicate that bridging chains are non-uniformly distributed throughout the sample, and pullout of the endblocks occurs over a range of applied stresses. 

\IRPS{For all $c$, the emulsions containing SEOS-180/32 display an inverse relationship between $\omega$ and $\beta$ with a minimal dependence on $c$. At low $\omega$, sufficient time is provided for chains to relax to accommodate the applied stress. Thus, most chains are at a similar stress state, resulting in sharp yielding in contrast to what occurs when the system is yielded rapidly. At high $\omega$, insufficient time results in chains existing at a range of stress states when chain pullout occurs, resulting in a relatively shallow slope.} This strong frequency dependence is consistent with the G'' overshoot data which showed greater energy dissipation at higher frequencies due to the inability of the chains to fully relax at high shear rates prior to chain pullout. Additionally, these findings indicating significant stress heterogeneity in the sample are consistent with simulations on colloidal gels.\cite{Colombo2014} 

Such sharp yielding and frequency dependence, however, are not observed in the emulsions containing SEOS-180/10. The sharpness of the yielding of the emulsions containing SEOS-180/10 -- which possesses a high-$\mathrm{M}_\mathrm{w}$ midblock and low-$\mathrm{M}_\mathrm{w}$ endblocks -- is relatively low for all $c$, but shows slight increases with increasing $c$. Combined with the small overshoots in G'' observed for SEOS-180/10, we conclude that there is little energy dissipation upon yielding in the emulsions containing this polymer, and the yielding tends to occur over a broad range of stresses. Finally, we consider the emulsions containing SEOS-27/11.5, which has low-$\mathrm{M}_\mathrm{w}$ mid- and endblocks. At the lowest concentration of polymer, the emulsions are largely fluids, displaying low $\beta$ values regardless of $\omega$. The emulsions containing higher concentrations of SEOS-27/11.5, however, display a similar inverse relationship between $\beta$ and $\omega$ as those containing SEOS-180/32. This frequency dependence suggests that the breaking of interdroplet connections occurs at a narrow distribution of stress states at low $\omega$ and at a broader distribution at high $\omega$. Therefore, yielding the material at a higher frequency causes endblock pullout at a broader range of stresses, while yielding the material at lower $\omega$ provides sufficient time for the system to equilibrate the tension on each bridging chain. This competition between endblock pullout and midblock stretching explains the observed frequency dependencies and differences in the energy dissipation upon yielding of these materials. 

Overall, these frequency effects suggest that the segmental relaxations of the midblock can play a significant role during yielding. At low $\omega$, segmental relaxations relieve tension on bridging chains and allow the system to take on a more energetically favorable structure. At high frequencies, however, the polymer midblocks have insufficient time to relax, resulting in the stresses being distributed heterogeneously across the network. Bridges therefore experience pullout at different stresses, resulting in a broader yielding and lower values of $\beta$. In summary, we find that these polymer-linked emulsions yield through droplet rearrangement, with the properties of the linking polymer determining the extent of energy dissipation and the sharpness of the yielding. 

\subsection{Yielding Effects on Structure}

\begin{figure*}[tb!]
\includegraphics[width = 6.5 in]{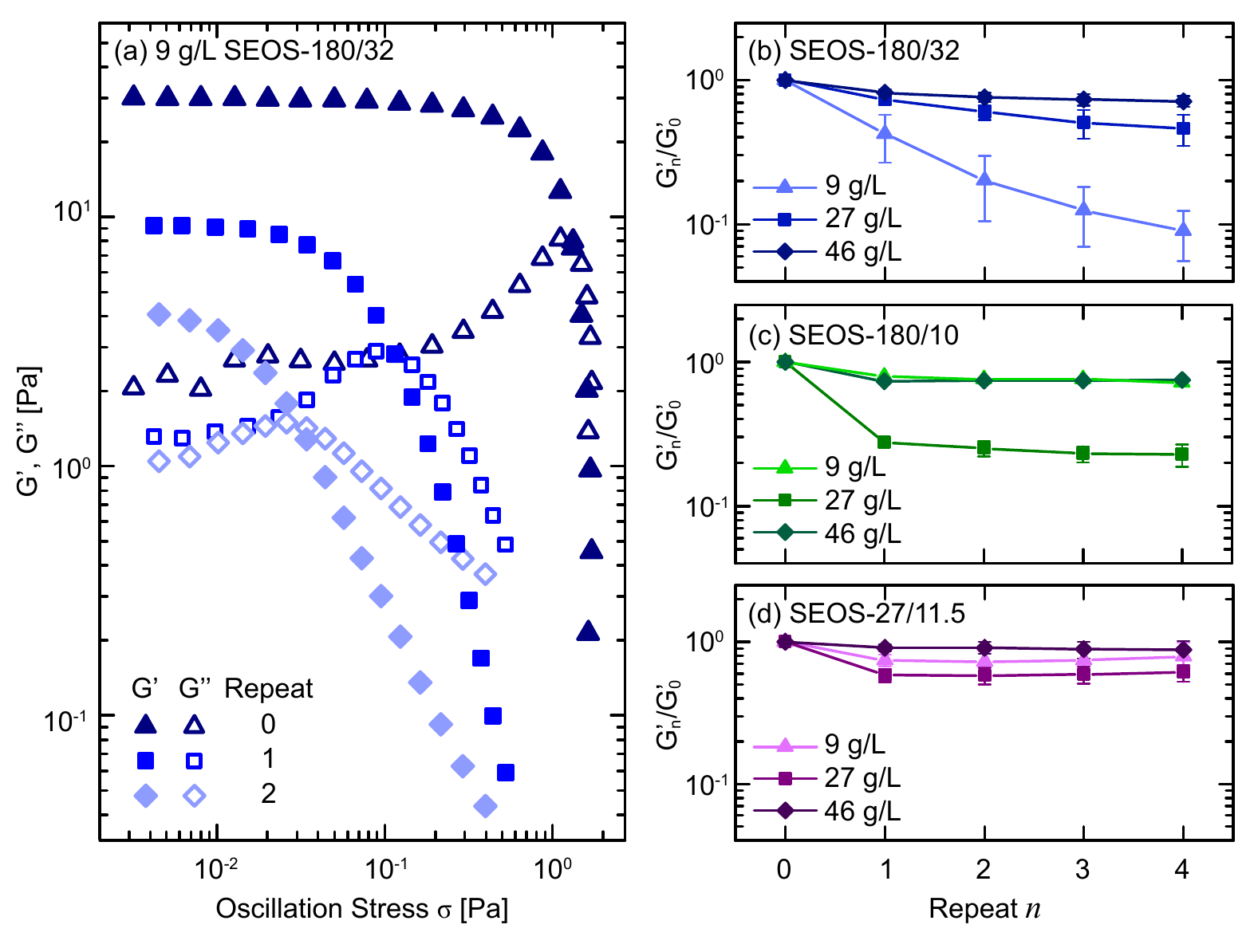}
\caption{\label{fig:storVSrun} (a) Storage modulus $\mathrm{G'}$ (closed) and loss modulus $\mathrm{G''}$ (open) as a function of amplitude stress $\sigma$ for an emulsion containing 9.0 g/L SEOS-180/32. (b-d) Normalized G'/G$_0$ as a function of run $n$ for emulsions containing the specified concentrations of (b) SEOS-180/32, (c) SEOS-180/10, and (d) SEOS-27/11.5. Points represent mean values and error bars show one standard deviations. }
\end{figure*}

We have now developed a physical picture of the mechanism by which these emulsions yield, with the $\mathrm{M}_\mathrm{w}$ of the linking polymers dictating the extent of energy dissipation and the sharpness of the yielding process, but how the emulsion structure is altered upon yielding remains to be explored. Yielding these linked emulsions forces the endblocks to pull out of the emulsion droplets before rapidly re-embedding in an emulsion droplet to reform a bridge or to generate a loop. We therefore expect the bridging density to decrease after yielding, as some bridges are converted to loops during network rearrangement. We test this hypothesis by subjecting the emulsions to repeated amplitude sweeps and evaluating changes to the material strength and energy dissipation (Fig. \ref{fig:storVSrun}). For emulsions containing SEOS-180/32, the linear elasticity, the height of the G'' overshoot, and the sharpness of the yield transition all decrease with each additional amplitude sweep conducted. This progressive decrease in the material elasticity suggests that the network weakens under repeated yielding processes as more chains are pulled out and converted to loops. Moreover, as fewer bridges remain, yielding the material disrupts fewer chains to dissipate energy, resulting in the diminishing G'' overshoot height. This continuous decrease in material strength is observed for all emulsions containing SEOS-180/32, with emulsions containing lower concentrations displaying more significant weakening of the network (Fig. \ref{fig:storVSrun}b). Emulsions containing only 9 g/L of SEOS-180/32 display a decrease in G' of an order of magnitude after four yieldings, whereas those containing higher concentrations only lose 10-50\% of their initial strength over the same period. Regardless of concentration, the rate of network strength weakening decreases each time the emulsions are yielded, indicating that fewer and fewer bridges are broken.

A very different trend is observed for the emulsions containing lower $\mathrm{M}_\mathrm{w}$ SEOS (Fig. \ref{fig:storVSrun}c-d). Although all samples show a decrease in $\textrm{G'}_\textrm{n}/\textrm{G'}_\textrm{0}$ from the initial run to the first repeat, the samples containing SEOS-180/10 or SEOS-27/11.5 show little change with repeated amplitude sweeps beyond the first. Instead, G' plateaus and becomes independent of the number of yielding events the material has undergone. This plateau indicates that minimal chain pullout occurs for these emulsions after the first yielding event. It is notable that the emulsions do remain elastic, however, with linear moduli greater than that of the neat emulsions or emulsions containing PEO. We theorize that these emulsions linked by low-M$_\mathrm{w}$ polymers possess a structure consisting of weakly connected clusters of linked droplets.\cite{Manley2005,Lu2006} The networks of linked clusters may initially percolate the sample, but they break apart into the constituent clusters upon the application of high shear, resulting in a fluid of clusters. Upon further shearing of the emulsion, the clusters can undergo rearrangement instead of further network breakdown, so the moduli becomes independent of the number of additional amplitude sweeps. This physical picture contrasts with the structure of the emulsions containing SEOS-180/32, which are expected to possess densely connected, percolated structures.

\begin{figure*}[tb!]
\includegraphics[width = 6.5 in]{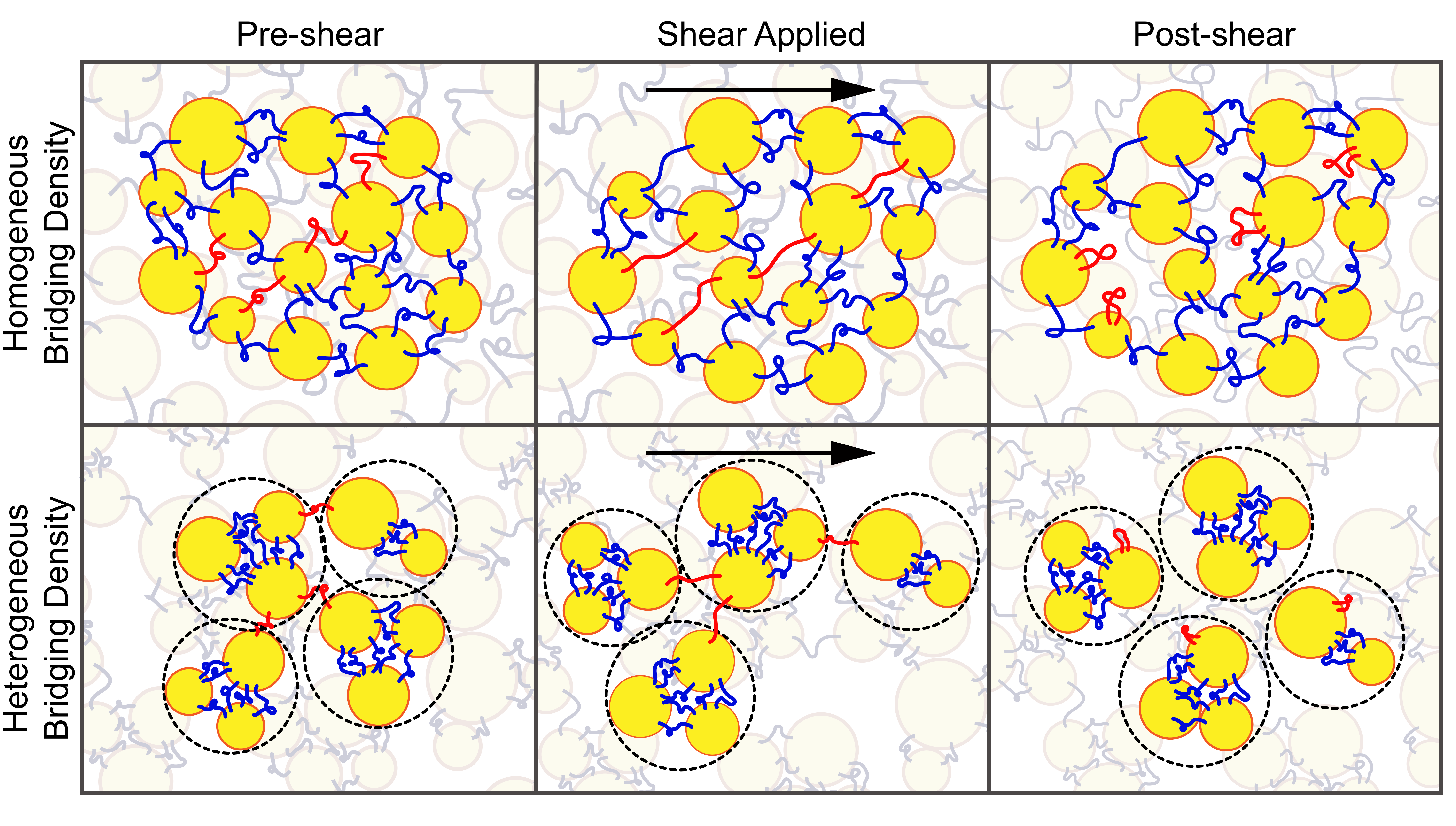}
\caption{\label{fig:underShear} Schematic depicting the effect of shear on systems with homogeneous or heterogeneous bridging density. The schematic depicts the emulsion prior to shear being applied, under shear, and after shearing is complete. Chains that transition from bridging to looping are shown in red. }
\end{figure*}

We rationalize that these different microstructures form as a consequence of varying degrees of bridging heterogeneity in the emulsions, which result from kinetic trapping and size-dependent bridging probabilities. The polymers with high-$\mathrm{M}_\mathrm{w}$ midblocks easily span between droplets without experiencing a large entropic penalty, meaning that the polymer can form bridges throughout the material. The polymers with low-$\mathrm{M}_\mathrm{w}$ midblocks, however, can only form bridges when the particles are already in close proximity. During the initial dissolution of low-M$_\mathrm{w}$ polymer, bridges between droplets are more likely to be reinforced with additional bridges due to the close proximity of bridged droplets. SEOS-180/10, with a high-$\mathrm{M}_\mathrm{w}$ midblock and low-$\mathrm{M}_\mathrm{w}$ endblock appear to possess a similar structure, but in this case we expect that the weak partitioning of the endblocks allows the chains to rearrange and concentrate in regions where droplets are closely packed. A schematic of these two different structures is shown in Fig. \ref{fig:underShear}. The difference in probability of bridge formation between two already-bridged droplets versus two unconnected droplets results in clusters with high internal bridging density. When the networks contain a relatively homogeneous distribution of bridges, as shown in the top half of Fig. \ref{fig:underShear}, the application of high shear will result in some bridges in the network converting to loops, but the overall percolated structure will remain. This is in contrast to a system with a heterogeneous bridging density, as shown in the bottom half of the schematic. In this case, the application of shear will break weak links between more heavily bridged sections of the network, resulting in the network breaking apart producing a fluid of clusters.

\subsection{Confirmation of Structures}
\begin{figure}[tb!]
\includegraphics[width = 3.25 in]{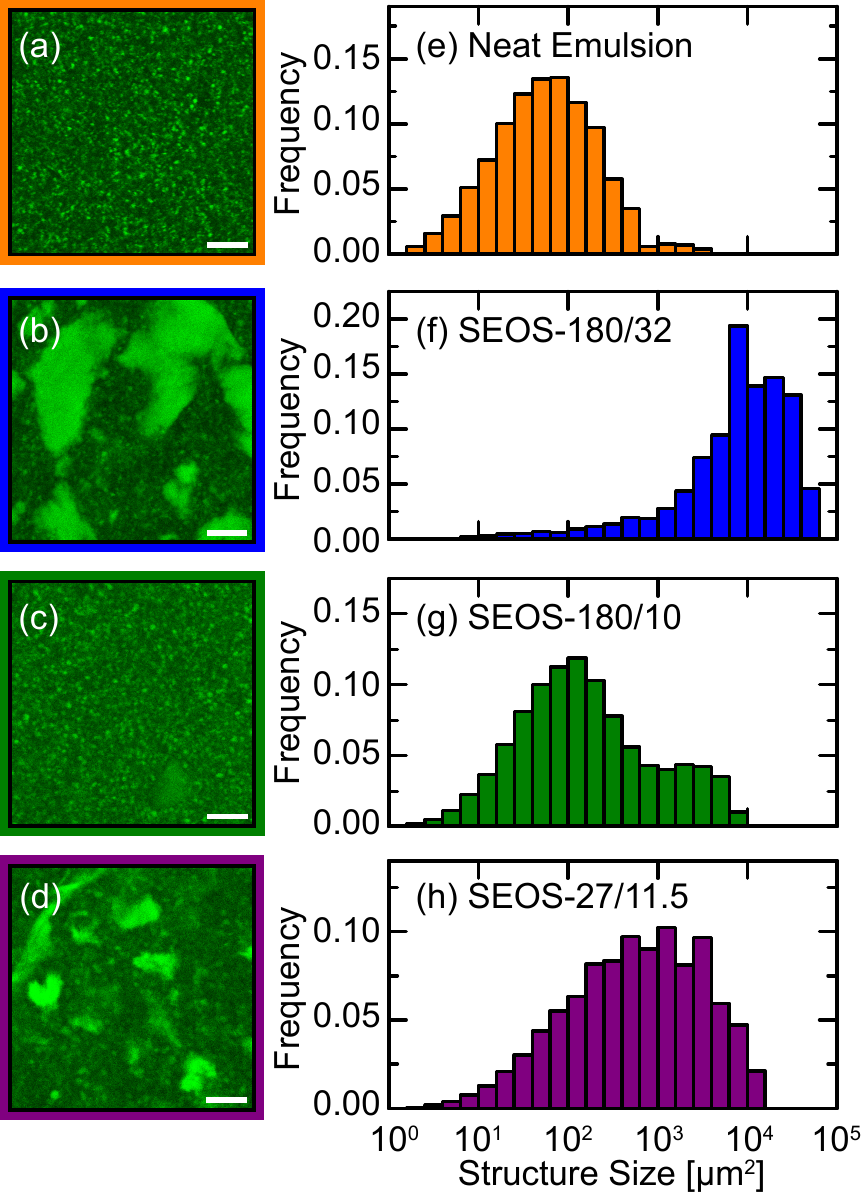}
\caption{\label{fig:clusters} (a-d) Confocal images of index-matched (a) emulsions containing (a) no polymer, (b) SEOS-180/32, (c) SEOS-180/10, and (d) SEOS-27/11.5. $c=9$ g/L for polymer containing emulsions and $\phi=0.5$ prior to dilution. The final diluted emulsions contain approximately 4.3 g/L and $\phi=0.24$. Scale bars are 50 $\mu$m in length. (e-h) Histograms presenting weighted frequencies of cluster sizes in (e) neat emulsions and emulsions containing (f) SEOS-180/32, (g) SEOS-180/10, and (h) SEOS-27/11.5.}
\end{figure}

The nonlinear rheology of these emulsions suggests that polymer $\mathrm{M}_\mathrm{w}$ controls the structure of the elastic network with higher $\mathrm{M}_\mathrm{w}$ polymers forming a homogeneous percolated structure and lower $\mathrm{M}_\mathrm{w}$ polymers inducing the formation of weakly linked clusters of droplets. We confirm the presence of these different structures through confocal microscopy of index-matched emulsions. Emulsions are diluted with glycerol to bring the continuous phase to 68.5\% glycerol such that the continuous phase has the same refractive index as the dispersed cyclohexane. The final polymer concentration and volume fraction after dilution are $c=4.3$ g/L and $\phi=0.24$, respectively. \EDITSDK{We adopt this dilution protocol for two reasons. First, emulsions prepared at the index-matched solvent composition exhibit significant coarsening, which we attribute to poor dissolution of the triblock copolymer in the glycerol solution. Second, we can not distinguish between connected and dispersed droplets at high volume fractions (i.e., $\phi = 0.5$) because there is no observable difference in particle separation and because the polymer is not visible in confocal microscopy.} Under confocal microscopy, we observe stark differences in the degree of clustering and the size of the clusters \EDITSDK{for these diluted samples} (Fig. \ref{fig:clusters}a-d). Although the dilution step may disrupt the local structure, multiple emulsion were imaged with confocal microscopy and showed quantitatively similar features, indicating that these measurements are reproducible. 

The neat emulsions exhibit a homogeneous well-dispersed structure. In these micrographs, each feature contains multiple droplets which cannot be individually distinguished due to the resolution of the image and particle motion. In the presence of the telechelic polymer though, droplets form larger fractal structures dispersed in a suspension of individual droplets. Emulsions containing SEOS-180/32 show a small population of very large features, which cover a significant portion of the image surface. The emulsions containing SEOS-180/10 or SEOS-27/11.5 also display clusters, but these clusters are smaller than those observed in the emulsions containing the larger $\mathrm{M}_\mathrm{w}$ endblock polymer. Additionally, we observe fewer large features in the emulsion containing SEOS-180/10 than in those containing SEOS-27/11.5, suggesting a difference in how these two polymers induce droplet aggregation. We quantify these structural differences according to the area-weighted probability distributions shown in the histograms of Fig. \ref{fig:clusters}e-h. The weighted distribution of structures in the neat emulsions peaks under 100 $\mathrm{\mu m}^2$, which serves as a control to which we can compare the emulsions containing SEOS. Error is introduced into this quantification due to both the image resolution and the algorithm's binarization step. \EDITSDK{The confocal images are acquired over a 20 s interval due to signal limitations, during which a 400 nm droplet is expected to diffuse approximately 1.4 $\mu$m based on Stokes-Einstein predictions and a viscosity of 0.0225 Pa$\cdot$s for the glycerol water mixture. This diffusive motion is larger than the average interparticle spacing of approximately 550 nm for droplets of this size and volume fraction, resulting in an overlap of individual point spread functions. As a result,} droplets in close proximity may appear linked, resulting in anomalously large features ($\mathcal{O}$(100 $\mu$m$^{2}$)) being identified even in the neat emulsions.

The weighted distribution of structure sizes in the emulsions containing SEOS-180/32 is dominated by the presence of very large structures, ranging from $10^3$-$10^5$ $\mathrm{\mu m}^2$ in size. The neat emulsions show no features this large, indicating that these large structures are clusters of polymer-linked droplets. \EDITSDK{Although individual droplets cannot be resolved within the cluster, we know that these structures do not arise from liquid-liquid or gel phase separation because of their non-spherical shape and from the fact that the polymer is insoluble in either indivudal solvent. Furthermore, additional dilution results in the emergence of two populations corresponding to individual droplets and small clusters, as confirmed by dynamic light scattering\cite{Keane2022}}. In the emulsions containing SEOS-180/10, we observe many individual droplets, similar to the features identified in the neat emulsion, with the addition of a small podpulation of structures in the size range of $10^3$-$10^4$ $\mathrm{\mu m}^2$. In the emulsions containing SEOS-27/11.5, we observe a greater number of structures in the size range of $10^3$-$10^4$ $\mathrm{\mu m}^2$. The droplet clusters observed in the emulsions containing SEOS-180/32 are significantly larger than those identified in the emulsions containing the lower $\mathrm{M}_\mathrm{w}$ polymers indicating that the SEOS-180/32 forms a more homogeneous percolating network. 

These results are consistent with the structures we hypothesized based off of the nonlinear rheology and that are shown schematically in Fig. \ref{fig:underShear}. The linked emulsions consist of areas of high and low bridging density, with the emulsions containing SEOS-180/32 possessing large structures that percolate the sample. Within individual clusters, the bridging is dense enough to develop strong networks, but weaker regions are likely broken in the process of dilution. For the emulsions containing small clusters, dilution breaks the network in areas of low bridging density resulting in a fluid of small clusters. This bridge-breaking is akin to the network breakdown which occurs under shear, producing clusters that can rearrange without further breaking of bridges.

\begin{figure*}[tb!]
\includegraphics[width = 6.5 in]{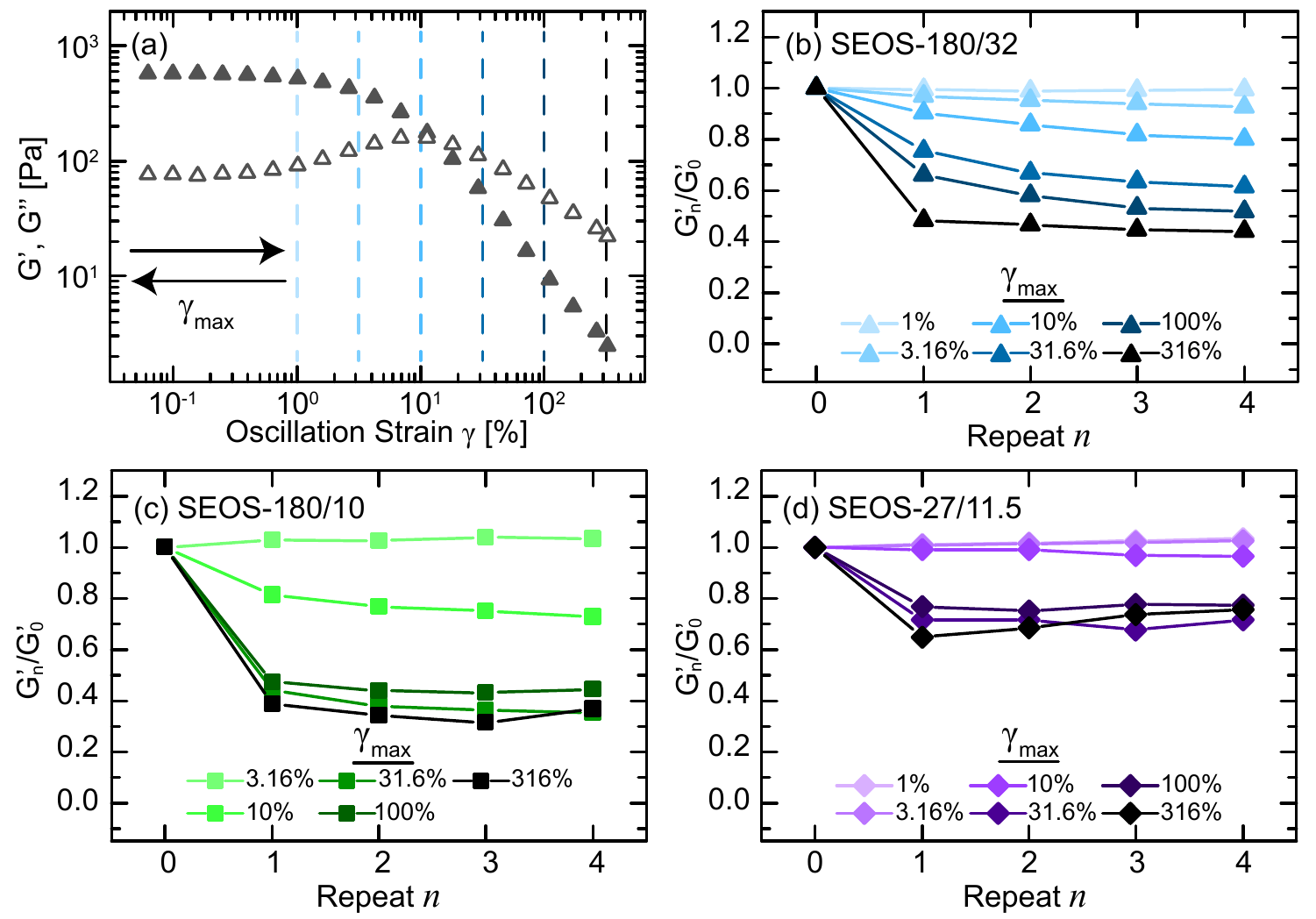}
\caption{\label{fig:diffStrains} A visual to demonstrate the procedure used to collect data for plots \ref{fig:diffStrains}b-d, showing repeated amplitude sweeps up to a maximum oscillation strain $\gamma_\mathrm{max}$. (b)-(d) The storage modulus G'$_\mathrm{LVR}$ in the LVR normalized by G'$_\mathrm{LVR}$ in the initial run as a function of the number of repeated amplitude sweeps, varying the maximum oscillation strain $\gamma_\mathrm{max}$. The emulsions contain 27 g/L of (b) SEOS-180/32, (c) SEOS-180/10, and (d) SEOS-27/11.5. All measurements are taken with $\omega=$10 rad/s.}
\end{figure*}

As a final test of our physical picture, we conduct repeated amplitude sweeps varying the maximum strain $\gamma_\mathrm{max}$, as shown in Fig. \ref{fig:diffStrains}. The strains are chosen to be logarithmically spaced and span from the LVR to the maximum strain limit of the rheometer. We again observe distinct differences between the emulsions containing SEOS-180/32 and those containing the other $\mathrm{M}_\textrm{w}$ copolymers. For all emulsions, amplitude sweeps conducted fully within the LVR do not appreciably decrease the emulsion elasticity as the strain is too small to force the endblocks to pull out. When $\gamma_\mathrm{max}$ exceeds the LVR, however, G' decreases with repeat number. For the emulsions containing SEOS-180/32, the rate of decay with repeated amplitude sweeps depends on $\gamma_\mathrm{max}$, with larger strains resulting in faster decays. This more rapid yielding suggests that greater strain causes more bridges to break throughout the percolated network. In contrast, the emulsions containing either SEOS-180/10 or SEOS-27/11.5 exhibit discontinuous behaviors across a critical strain amplitude between 10\% and 31.6\%. When yielded at strains above this value, the emulsion elasticity precipitously drops during the first yielding event and then plateaus. This behavior indicates that exceeding a certain applied stress causes endblock pullout at areas of low bridging density, producing a fluid of clusters. As the material is further sheared, these clusters simply rearrange instead of further endblock pullout in the areas of higher bridging density. \IRPS{From these measurements, we demonstrate how emulsion structure and bridging heterogeneity modify complex processing-property relationships for polymer-linked emulsions. We expect that additional control over these properties may be achieved by conducting similar experiments at varying frequencies to evaluate the role of segmental relaxations of the linking polymer, which we aim to investigate in future work.}

\section{Conclusions}
In this work, we explore the nonlinear rheology of gels formed from the physical crosslinking of cyclohexane-in-water emulsions by a telechelic, triblock copolymer polystyrene-\textit{b}-poly(ethylene oxide)-\textit{b}-polystyrene (SEOS). We show that these polymer-linked emulsions yield primarily through the rearrangement of dispersed droplets, similar to behavior of other amorphous materials, but now facilitated by the relaxations and pullout of the bridged polymer. Depending on the $\mathrm{M}_\mathrm{w}$ of the block copolymer, we observe stark differences in the yielding behavior of the emulsions, including the extent of energy dissipation, the sharpness of the yield transition, and the response to processing. For all polymers, increasing the concentration $c$ leads to a more well-connected network with higher bridging density. The distribution of these bridges in the system, however, depends on the $\mathrm{M}_\mathrm{w}$ of the linking polymer's constituent blocks, resulting in significantly different yielding behavior and responses to processing. The emulsions containing high-$\mathrm{M}_\mathrm{w}$ linkers tend to take on the structure of a percolated network, whereas those containing low-$\mathrm{M}_\mathrm{w}$ linkers are generally composed of weakly-linked clusters of droplets. 

Although yield stress fluids are ubiquitous throughout everyday life and across industries, it remains challenging to design materials with controlled yield stresses. Our work demonstrates a unique approach by which the yield stress and mechanism can be programmed into complex fluids by exploiting the distinct chemistry of triblock copolymers. The $\mathrm{M}_\mathrm{w}$ of the constituent blocks can be varied to control the structure of the dispersed phase, resulting in materials with significantly different yielding responses. We expect these findings will enable the development of novel soft materials for applications ranging from 3D printing\cite{Sommer2017} to tissue scaffolds\cite{Lin2013,Aguilar-De-leyva2020}.

\begin{acknowledgement}
We would like to thank Dr. Simon Rogers for fruitful discussions. This research was supported in part by ACS Petroleum Research Fund Doctoral New Investigator grant 65826-DNI9 and in part by the Rhode Island Institutional Development Award (IDeA) Network of Biomedical Research Excellence from the National Institute of General Medical Sciences of the National Institutes of Health under grant number P20GM103430. 
\end{acknowledgement}

\begin{suppinfo}
The supporting information includes the molecular structure of polystyrene-\textit{b}-poly(ethylene oxide)-\textit{b}-polystyrene (SEOS), \EDITSDK{a schematic of the polymer in looping and bridging conformations,} a time sweep demonstrating our control over evaporation, visuals of the aqueous, index-matched, and dyed index matched emulsions, Lissajous-Bowditch figures for different samples and the associated Fourier transformations, \EDITSDK{Lissajous-Bowditch figures corresponding to the 46 g/L samples associated with Figure \ref{fig:slopesAndHeights}}, a schematic identifying yielding parameters, and moduli as a function of repeated amplitude sweeps for emulsions containing SEOS-180/10 and SEOS-27/11.5.
\end{suppinfo}

\bibliography{biblio}

\end{document}